\pdfoutput=1

\documentclass[prd,amsmath,amssymb,nofootinbib,preprintnumbers]{revtex4}

\usepackage{graphicx}
\graphicspath{{figures/}}

\def\lsim{\mathrel{\raise.3ex\hbox{$<$\kern-.75em\lower1ex\hbox{$\sim$}}}}
\def\gsim{\mathrel{\raise.3ex\hbox{$>$\kern-.75em\lower1ex\hbox{$\sim$}}}}

\newcommand{\like}{\ensuremath{\mathcal{L}}}
\newcommand{\mchi}{\ensuremath{m_{\chi}}}
\newcommand{\sigmav}{\ensuremath{\langle\sigma v\rangle}}
\newcommand{\hatsigmav}{\ensuremath{\widehat{\langle\sigma v\rangle}}}
\newcommand{\dNdE}{\ensuremath{\frac{dN}{dE}}}
\newcommand{\dphidE}{\ensuremath{\frac{d\phi}{dE}}}
\newcommand{\dphikdE}{\ensuremath{\frac{d\phi_{\makebox[0pt][l]{\tiny k}}}{dE}\phantom{i}}}
\newcommand{\dPhippdE}{\ensuremath{\frac{d\Phi_{\makebox[0pt][l]{\tiny pp}}}{dE}\phantom{i}}}
\newcommand{\hatJ}{\ensuremath{\widehat{\mathbf{J}}}}
\newcommand{\dhatJ}{\ensuremath{\widehat{\vphantom{\rule{0pt}{1.95ex}}\smash{\widehat{\mathbf{J}}}}}}

\begin{document}

\title{Fermi/LAT observations of Dwarf Galaxies highly constrain a Dark Matter Interpretation of Excess Positrons seen in AMS-02, HEAT, and PAMELA }

\author{Alejandro L\'{o}pez}
\email{aolopez@umich.edu}
\affiliation{Michigan Center for Theoretical Physics, University of Michigan, Ann Arbor, Michigan 48109-1040, USA}

\author{Christopher Savage}
\email{chris@savage.name}
\affiliation{Nordita (Nordic Institute for Theoretical Physics), KTH Royal Institute of Technology and Stockholm University, Roslagstullsbacken 23, SE-106~91 Stockholm, Sweden}

\author{Douglas Spolyar}
\email{dspolyar@gmail.com}
\affiliation{Oskar Klein Centre for Cosmoparticle Physics, Stockholm University, Stockholm, Sweden}

\author{Douglas Q.\ Adams}
\email{doug.q.adams@gmail.com}
\affiliation{{}99~Battery Place, New York, NY 10280, USA}

\date{\today}

\preprint{NORDITA-2015-005, MCTP-15-03}

\begin{abstract}

It is shown that a Weakly Interacting Massive dark matter Particle
(WIMP) interpretation for the positron excess observed in a variety of
experiments, HEAT, PAMELA, and AMS-02, is highly constrained by the
Fermi/LAT observations of dwarf galaxies.  In particular, this paper
has focused on the annihilation channels that best fit the current
AMS-02 data (Boudaud et~al., 2014).  The Fermi satellite has surveyed
the $\gamma$-ray sky, and its observations of dwarf satellites are
used to place strong bounds on the annihilation of WIMPs into a
variety of channels.  For the single channel case, we find that dark
matter annihilation into
$\{b\bar{b},e^+e^-,\mu^+\mu^-,\tau^+\tau^-,4$-$e,$ or $4$-$\tau \}$ is
ruled out as an explanation of the AMS positron excess (here $b$
quarks are a proxy for all quarks, gauge and Higgs bosons).  In
addition, we find that the Fermi/LAT 2$\sigma$ upper limits, assuming
the best-fit AMS-02 branching ratios, exclude multichannel
combinations into $b\bar{b}$ and leptons.  The tension between the
results might relax if the branching ratios are allowed to deviate
from their best-fit values, though a substantial change would be
required.  Of all the channels we considered, the only viable channel
that survives the Fermi/LAT constraint and produces a good fit to the
AMS-02 data is annihilation (via a mediator) to 4-$\mu$, or mainly to
4-$\mu$ in the case of multichannel combinations.

\end{abstract}


\maketitle

Weakly Interacting Massive Particles (WIMPS), such as the lightest
supersymmetric particles (for reviews, see
Refs.~\cite{Jungman:1995df,Bertone:2004pz}), are thought to be the
best motivated dark matter (DM) candidates.  The particles in
consideration are their own antiparticles; thus, they annihilate among
themselves in the early universe and naturally provide the correct
relic density today to explain the dark matter of the universe.  This
same annihilation process takes place in the present universe wherever
the DM density is sufficiently high and is the basis for DM indirect
detection searches. Indirect detection experiments search for the
annihilation products of dark matter particles, including electrons
and/or positrons, antiprotons, photons, and neutrinos. Promising sites
for the observation of dark matter annihilation products include the
core of the Sun \cite{Srednicki:1986vj}, the
Earth~\cite{Freese:1985qw,Krauss:1985aaa}, our Galactic
halo~\cite{Ellis:1988qp,Turner:1989kg,Kamionkowski:1990ty,Silk:1984zy},
Galactic center~\cite{Bergstrom:1997fj}, dwarf satellite
galaxies~\cite{Evans:2003sc,Bergstrom:2005qk}, and from DM
substructures~\cite{Bertone:2005xz,Sandick:2010yd,Sandick:2011zs,Lavalle:2012ef}.

Over the past several years, there have been a number of experimental
signals which have been interpreted as possible indications of DM.
Confirmation that any of these observations are actually due to DM,
rather than being a mere experimental artifact or astrophysical
background, would likely require more than one experiment to provide
complementary information.  In this paper, we consider the anomalous
features in the spectrum of cosmic ray positrons and electrons
reported by AMS-02 \cite{Accardo:2014lma},
PAMELA~\cite{Adriani:2008zr,Adriani:2013uda}, and the Large Area
Telescope of the Fermi Gamma Ray Space Telescope
(Fermi/LAT)~\cite{Collaboration:2009zk} (as well as in earlier
indications from
HEAT~\cite{Barwick:1997ig,DuVernois:2001bb,Beatty:2004cy}). The
positron fraction was found to be a steadily increasing function of
energy, above 10~GeV. This behavior is difficult to explain with
standard astrophysical mechanisms, in which positrons are secondary
particles, produced in the interactions of primary cosmic rays during
the propagation in the interstellar medium. These observations have
led to a great deal of speculation that DM annihilations~
\cite{Baltz:2001ir,Barger:2008su,Harnik:2008uu,Cirelli:2008pk,Nelson:2008hj,Cholis:2008qq,Zurek:2008qg,Fox:2008kb,Chen:2008dh,Cholis:2008wq,Hooper:2009fj,Cholis:2013lwa,Bergstrom:2013jra,Cholis:2013psa,Hooper:2009gm,Hooper:2004xn,Hooper:2003ad,Ibarra:2013zia,Cholis:2008hb,Dev:2013hka}
or decays~ \cite{Arvanitaki:2008hq,Nardi:2008ix,Arvanitaki:2009yb,Feng:2013zca} may
be responsible. However, any explanation of these positron/electron
signals in terms of DM annihilation requires somewhat nonstandard WIMP
properties.  In particular, the local halo density of DM within the
vicinity of the Solar System is insufficient to produce these
observations unless the annihilation cross section is considerably
larger than that typically expected for a thermal relic, or the
annihilation rate is otherwise supplemented by a large boost factor
$\sim 10^1 - 10^4$.  Such an enhancement could arise due to
astrophysics; for example, due to substructures in the DM
distribution; yet Ref.~\cite{Brun:2009aj} argues that the probability
of such a nearby DM clump is $<1$\%.  In this paper, we will instead
focus on the possibility of an annihilation cross-section that is
enhanced by the required boost factor compared to the standard thermal
annihilation.  Furthermore, only a handful of possibilities can
explain the spectral shape reported by AMS-02 and PAMELA, as well as
avoid overproducing antiprotons (in excess of what is observed)
\cite{Adriani:2008zq,Donato:2008jk,Adriani:2010rc,Adriani:2012paa}.
The most well-studied approach to satisfying all these constraints has
been leptophilic DM, i.e.\ the DM annihilations proceed largely to
leptons ($\mu^+\mu^-$ or $\tau^-\tau^-$), which do not produce any
antiprotons.  Alternatively, other annihilations channels---quarks,
vector and Higgs bosons---are allowed if the DM particle is heavier
than $\sim$10~TeV, producing antiprotons at higher energies than those
probed by AMS-02 data \cite{Cirelli:2008pk}.

Other explanations of the Cosmic Ray Positron Excess (henceforth CRPE)
have also been proposed. The most plausible is that it is due to
pulsars
\cite{Hooper:2008kg,Profumo:2008ms,Linden:2013mqa,Delahaye:2014osa}.

Positrons can ultimately be produced by DM annihilating into a variety
of channels: single channels (leptons, quarks, gauge bosons, Higgs, or
four leptons) or a mix of these channels.  Prior to 2014, when the
most recent AMS-02 results became available, all annihilation channels
could provide explanations for the positron excess (see Table~2 in
Ref.~\cite{Boudaud:2014dta}).  The most popularly studied cases were
WIMP masses $\sim$200~GeV, which matched the data only for leptophilic
channels (DM annihilates only into leptons $\mu^+ \mu^-$ or
$\tau^+ \tau^-$), though quark and gauge boson channels well fit the
data at higher WIMP masses, with best-fit masses as high as
$\sim$50--200~TeV (depending on the channel).

The recent AMS-02 data release \cite{Accardo:2014lma} has greatly
improved our understanding of the positron excess. As stressed by
Ref.~\cite{Boudaud:2014dta}, two major improvements have emerged.
First the new data are far more accurate and extend out to 500~GeV,
much higher energies than previously explored.  Improved accuracy in
the positron spectrum leads to stronger constraints on any model for
the origin of the positrons.  Second, AMS-02 has measured directly the
total electron and positron flux, the denominator in the positron
fraction, reducing systematic errors and leading to differences in
best-fit regions for the DM mass and cross section.
Ref.~\cite{Boudaud:2014dta} further stressed the importance of
addressing uncertainties in the cosmic ray (CR) propagation model, as
these models are of critical importance in assessing the true nature
of the positron excess.

Boudaud et~al.~\cite{Boudaud:2014dta} have performed an analysis with
the new AMS-02 data, exploring the DM models that could explain the
CRPE.  Compared to earlier data and analyses, the recent AMS-02 data
do indeed lead to much stronger constraints on suitable DM candidates
that can explain the $e^+$ excess and the best-fit masses have shifted
by a fair amount.  Primarily these authors used the benchmark set of
CR propagation model parameters known as MED in obtaining their
results; this is the model that best fits the B/C ratio in the cosmic
rays.  With the recent, more accurate data and using MED, these
authors found that many of the earlier allowed models have been ruled
out (see Table~1 in Ref.~\cite{Boudaud:2014dta}).  In terms of single
channel models, leptophillic DM is no longer viable, whereas single
channel annihilation into quarks and gauge bosons provide an excellent
fit to the CRPE.  They also considered the case of DM annihilation
into a mediator particle $\phi$, where $\chi\chi\to\phi\phi$ with
$\phi\to\ell\bar{\ell}$, thus yielding four leptons per annihilation
(``4-lepton'' channel).  Again, single channel annihilation into four
leptons does not fit the AMS-02 data (assuming MED propagation).
Alternatively, annihilation into a combination of channels can also
provide an excellent fit, in particular annihilation into an admixture
of leptons and $b\bar{b}$ pairs.  Finally, a combination of four
lepton channels (arising from a mediator field $\phi$), specifically
the four-tau (75\%) and four-electron (25\%) channel, turns out to
provide a good fit to the AMS-02 data for a DM mass between 0.5 and
1~TeV.  The results reprised in this paragraph all refer to the MED CR
propagation model.

Far larger than the statistical errors in the AMS-02 data are the
systematic errors associated with the CR propagation.  Thus Boudaud
et~al.~\cite{Boudaud:2014dta} looked at a set of 1623 different
combinations of the CR transportation parameters---all
consistent with observed boron-to-carbon ratios---which bracket the
systematic uncertainty in the propagation model (see also
Ref.~\cite{Lin:2014vja} for an analysis of the new AMS-02 data, with a
different treatment of the propagation).  By including these
propagation model parameters into the fits to the AMS-02 data,
specifically by finding the DM parameters that fit the data for any of
the 1623 propagation parameter combinations, the allowed DM parameter
space increases.  In terms of the single channel models, the 2-$\tau$,
4-$\mu$, and 4-$\tau$ cases can now provide excellent fits to the
positron excess in addition to the cases mentioned above.

DM annihilations that produce $e^+$, either directly or through decays
and showering of the primary annihilation products, will invariably
also produce $\gamma$-rays.  Thus we can constrain the above DM models
by comparing with $\gamma$-ray observations.  The best places to look
are regions with a large abundance of DM: the Galactic Center,
clusters, and dwarf galaxies.  We will focus on dwarf galaxies for the
remainder of the paper. The dwarf spheroidal galaxies inside the Milky
Way are some of the most dark matter dominated objects known, with
mass-to-light ratios as high as $\sim$ 1000.  Because they are so dark
matter rich and nearby, they are exceptionally good places to
indirectly detect DM via $\gamma$-rays produced in its annihilation.
For previous $\gamma$-ray data that could be used to constrain a DM
interpretation of a positron excess, see the previous Fermi/LAT
combined analysis of dwarf galaxies \cite{Ackermann:2013yva}, VERITAS
\cite{Aliu:2012ga}, and MAGIC \cite{Aleksic:2013xea} observations of
Segue~1, as well as results of the H.E.S.S.\ collaboration
\cite{Abramowski:2014tra} on Sagittarius and other dwarf
galaxies.

Fermi/LAT surveys the $\gamma$-ray sky and, specifically, has looked
for $\gamma$-rays from 25 dwarf galaxies, detecting no significant
excess \cite{Ackermann:2013yva}.  This lack of signal is used to place
$\gamma$-ray flux upper limits for energies between 500~MeV and
500~GeV.  These bounds are then used to constrain the dark matter
annihilation for a broad range of particle masses and annihilation
channels.

The differential $\gamma$-ray flux $\dphidE$ from DM annihilation in a
dwarf galaxy can be written as the product of two components, a factor
$\dPhippdE\ $ that encodes all the particle physics and the so-called
J-factor that contains the astrophysics,
\begin{equation} \label{eqn:dphidE}
  \dphidE = \dPhippdE \times J
          = \bigg(\frac{1}{8\pi}\frac{\langle \sigma v \rangle}{m_\chi^2}\,\dNdE\bigg)
            \times\bigg(\int_{\Delta\Omega}\int_{\textrm{l.o.s.}} \rho^2(r)\,dr\,d\Omega \bigg) \, .
\end{equation}
Here, $\sigmav$ is the DM annihilation cross section, $m_\chi$ is the
DM mass, and $\dNdE$ is the photon spectrum from the DM annihilation,
which depends on the DM mass $\mchi$ and the annihilation channel.
The J-factor (the term in the second set of parentheses) integrates
the square of the DM density $\rho_\chi$ along the line of sight and
over a solid angle $\Delta\Omega$.  The J-factor can be estimated from
stellar kinematics as stars act as tracers of the gravitational
potential, allowing the DM distribution to be inferred.

We use Fermi/LAT dwarf $\gamma$-ray results to constrain DM models
using the data and likelihood technique described by the Fermi/LAT
collaboration in Ref.~\cite{Ackermann:2013yva}, to be briefly reviewed
here.  The primary quantity used in the likelihood analysis is the
energy flux
\begin{equation} \label{eqn:signal}
  s_{k,j} = \int_{E_{j,\min}}^{E_{j,\max}} E\,\dphikdE\,dE
\end{equation}
for each dwarf (indexed by $k$) and energy bin (indexed by $j$).
Here, $\dphikdE\equiv\dPhippdE\times J_k$ is the differential flux for
a dwarf with J-factor $J_k$.  For each dwarf and energy bin, Fermi/LAT
provides a likelihood $\like_{k,j}$ in $s_{k,j}$. The likelihood
function accounts for instrument performance, the observed counts,
exposure, and background fluxes.  For a given annihilation channel,
the energy flux is dependent only on the theoretical parameters
$\mchi$, $\sigmav$, and $J_k$; i.e.~$s_{k,j} =
s_{k,j}(\mchi,\sigmav,J_k)$.

Accounting for observational constraints on the J-factor, the
likelihood for a given dwarf $\like_k$ is
\begin{equation} \label{eqn:dwarflike}
  \like_k(\mchi,\sigmav,J_k)
    = \mathcal{LN}(J_k|\bar{J}_k,\sigma_k)
      \prod_j \like_{k,j}\big(s_{k,j}(\mchi,\sigmav,J_k)\big) \, ,
\end{equation}
where $\mathcal{LN}$ represents a log-normal distribution and
$\bar{J}_k$ \& $\sigma_k$ are the parameters describing that
distribution, derived from the stellar kinematics in the dwarf.  The
combined likelihood for multiple dwarfs is
\begin{equation} \label{eqn:dwarfslike}
  \like(\mchi,\sigmav,\mathbf{J})
    = \prod_k \like_{k}(\mchi,\sigmav,J_k) \, ,
\end{equation}
where $\mathbf{J}$ represents the set of J-factors $\{J_k\}$, one for
each dwarf.  We use in our analysis the same 15 non-overlapping dwarf
galaxies with J-factor estimates that are used by Fermi/LAT in their
analysis.  We use the J-factor estimates based upon a
Navarro-Frank-White (NFW) density profile \cite{Navarro:1996gj},
though a Burkert profile \cite{Burkert:1995yz} would not significantly
affect our results.\footnote{
  References~\cite{Strigari:2007at,Ackermann:2013yva} have shown that
  the the integrated J-factor within 0.5 degrees is fairly insensitive
  to the choice of dark matter density profile so long as the central
  value of the slope is less than 1.2.}

Fermi/LAT constraints in $\sigmav$ at a given $\mchi$ are determined
using a delta-log-likelihood approach treating the J-factors as nuisance
parameters.  The delta-log-likelihood $\Delta\like$ is given by
\begin{equation} \label{eqn:deltaloglike}
  \Delta\ln\like(\mchi,\sigmav)
    \equiv \ln\like(\mchi,\sigmav,\dhatJ)
           - \ln\like(\mchi,\hatsigmav,\hatJ) \, ,
\end{equation}
where $\hatsigmav$ \& $\hatJ$ are the values of $\sigmav$ \&
$\mathbf{J}$ that jointly maximize the likelihood at the given $\mchi$
and $\dhatJ\equiv\dhatJ(\mchi,\sigmav)$ are the J-factors that
maximize the likelihood for the given $\mchi$ and $\sigmav$.  The 1D
confidence intervals in $\sigmav$ at the $n\sigma$ confidence level
are determined by identifying the range of $\sigmav$ such that
\begin{equation} \label{eqn:nsigma}
  \Delta\ln\like(\mchi,\sigmav) \le n^2/2 \, .
\end{equation}
We will generally show the upper limit of the 2$\sigma$ confidence
intervals (95.4\%~confidence level).

We consider the case of DM annihilating directly to $b\bar{b}$ and
leptons $\ell\bar{\ell}$, as well as the ``4-lepton'' channels (via a
mediator $\phi$).  We use the $b\bar{b}$ case as a proxy for all other
quarks, gauge bosons, and the Higgs boson.  The spectra for the $u$,
$d$, $c$, $s$, and $t$ quarks; the $W$, $Z$, and $g$ gauge bosons; and
the $H$ boson are all similar in shape to the $b$ spectrum and within
$\sim$50\% of the amplitude.  Such differences of at most a factor of
two in amplitude will not prove to be significant; thus the $b$
spectrum should be reasonably representative of these other cases.%
\footnote{
  Each of the AMS-02 and Fermi constraints vary by less than a factor of
  two from the $b\bar{b}$ case when looking at other quarks, but
  $b\bar{b}$ is ruled out by an order of magnitude, so a factor of two
  is not enough to evade the Fermi/LAT constraints.}
The spectra from leptons, on the other hand, depend on the flavor, so
we consider $e$, $\mu$, and $\tau$ separately.  The annihilation
spectra $\dNdE$ for these different channels are derived using
\texttt{PYTHIA8} \cite{Sjostrand:2007gs,Sjostrand:2006za}.  For the
first set of cases, annihilation directly to quarks or leptons, we
include final state radiation (FSR) in the \texttt{PYTHIA} simulations
as these are the primary source of photons for the $e^{+}e^{-}$ and
$\mu^{+}\mu^{-}$ channels; the FSR provides only a minor contribution
to the $b\bar{b}$ and $\tau^{+}\tau^{-}$ channels.  We do not include
FSR for the mediator case due to the complexity, though that is again
the primary source of photons for the 4-$e$ and 4-$\mu$ channels.

\begin{figure*}
  \includegraphics[keepaspectratio,width=0.49\textwidth]{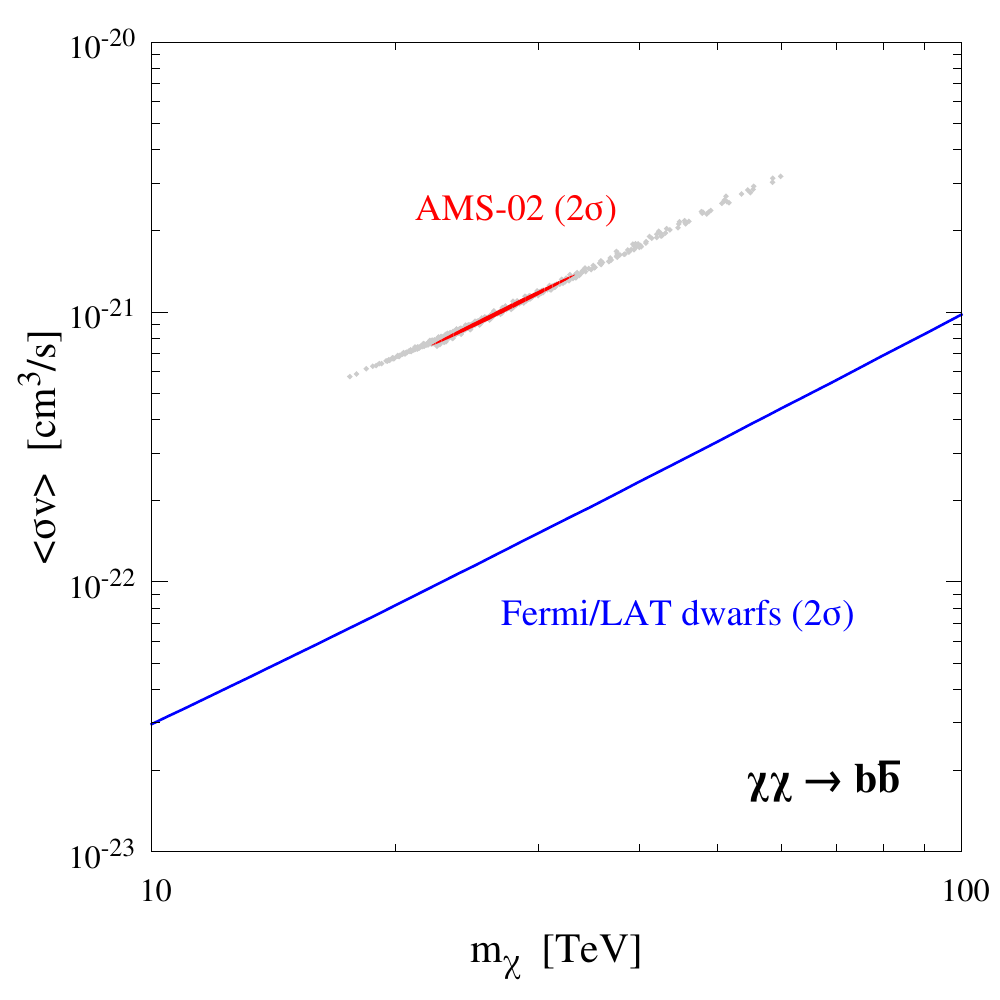}
  \hspace{\stretch{1}}
  \includegraphics[keepaspectratio,width=0.49\textwidth]{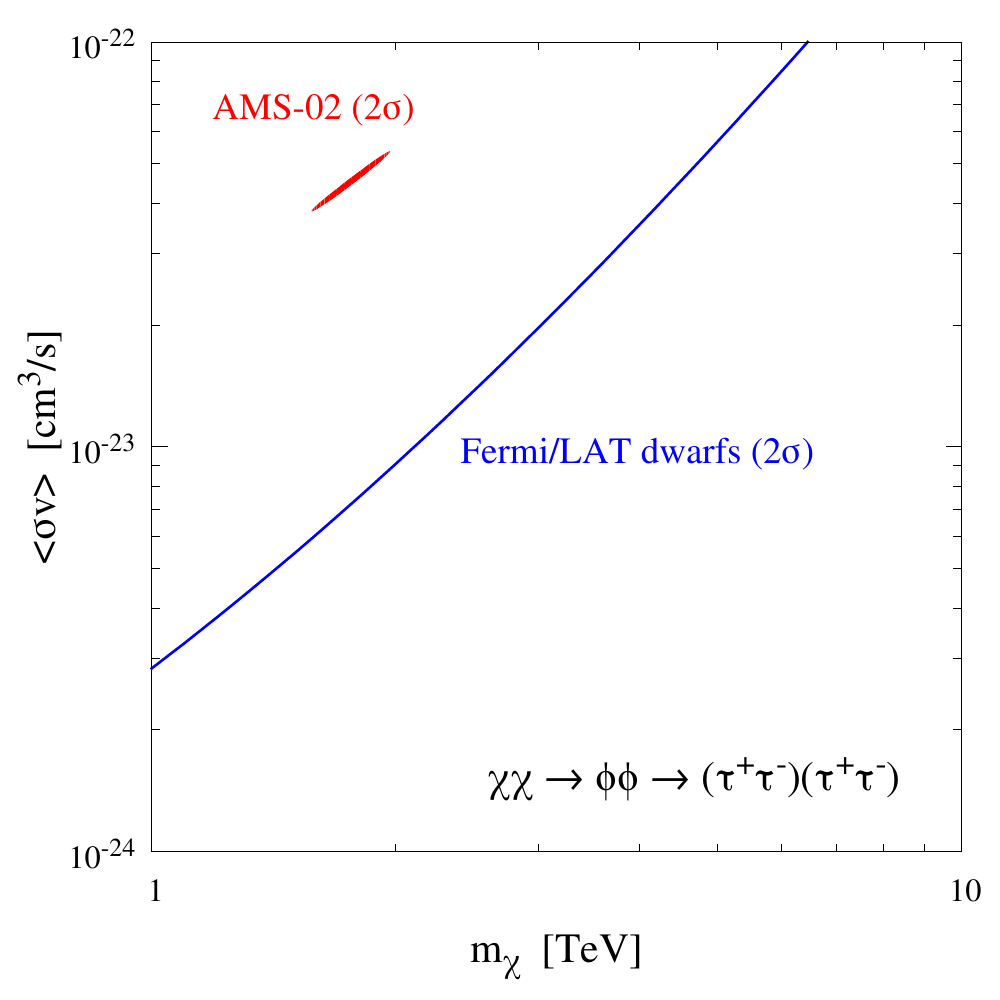}
  \caption{\label{fig:2b4tau}
    Constraints on the dark matter annihilation cross-section
    $\sigmav$ and mass $\mchi$ for annihilations into $b\bar{b}$
    (left) and four $\tau$'s via a mediator $\phi$ (right).  The
    best-fit AMS-02 parameters, as derived by
    Ref.~\cite{Boudaud:2014dta} for the MED propagation model parameters,
    are shown in red, while the Fermi/LAT upper bounds are shown by the
    blue curves.  Fermi/LAT dwarf constraints are generated using the
    procedure described in Ref.~\cite{Ackermann:2013yva}.  Constraints
    are shown at the 2$\sigma$ confidence level.  For the $b\bar{b}$
    case, the AMS-02 best-fit points for a selection of other
    cosmic ray (CR) propagation model parameters are shown as gray
    dots (also taken from Ref.~\cite{Boudaud:2014dta}).
  }
\end{figure*}

The AMS-02 2$\sigma$ confidence regions in $\sigmav$ vs.\ $\mchi$ for
the $b\bar{b}$ and 4-$\tau$ channels are shown in red in the left and
right panels, respectively, of Figure~\ref{fig:2b4tau}, taken from
Ref.~\cite{Boudaud:2014dta}.  The Fermi/LAT 2$\sigma$ upper limits in
$\sigmav$ are shown by the blue curves in the figure.  For both
channels, the AMS-02 regions are strongly excluded by the Fermi/LAT
dwarfs data.  In Figure~\ref{fig:2b4tau}, the AMS-02 2$\sigma$
confidence regions were determined assuming the MED propagation
parameters, the set of five astrophysical parameters fixed to best fit
the measured B/C ratio \cite{Donato:2003xg,Maurin:2001sj}.  As the
choice of propagation parameters has an impact on the fit to the
AMS-02 data, we show also the best-fit points found by
Ref.~\cite{Boudaud:2014dta} for a selection of other CR propagation
model parameters as discussed above.  These best-fit points, shown as
gray dots for the $b\bar{b}$ case, are taken from Figure~14 of
Ref.~\cite{Boudaud:2014dta} (parameters with $p\ge0.0455$ only).
Though varying the propagation parameters broadens the region of
parameter space consistent with the AMS-02 data, the DM interpretation
of the positron spectrum remains strongly in conflict with the
Fermi/LAT dwarf $\gamma$-ray observations.

We wish to comment on the differences between regions compatible with
AMS-02 assuming the MED propagation parameters and the regions
compatible when the propagation parameters are allowed to vary over
reasonable values, as represented by the 1623 sets of CR propagation
parameters.  In Figure~\ref{fig:2b4tau} one can see that all the
points, for both MED and other CR propagation model parameters, are in an
elongated region roughly parallel to the Fermi/LAT bounds.
Thus changing the CR propagation parameters does not alleviate the tension
between Fermi/LAT and a DM interpretation of AMS-02.  We believe that
this statement will be generally true, not just for the $b\bar{b}$
channel but for all channels (single and mixed).\footnote{%
  Since the annihilation rate scales as $\sigmav/\mchi^2$, moving
  along the elongation line corresponds to roughly a fixed number of
  $e^+$, as is required to match the AMS-02 data. There is a slight
  effect on the fit due to the $\mchi$ dependence of the spectrum,
  but the general trend still holds.}
Thus we believe the following: if a DM annihilation channel that fits
AMS-02 with MED propagation is ruled out by Fermi/LAT, then we believe
that the same channel is also likely to be ruled out if other
propagation parameters are used.  In other words, we do not believe
that using different reasonable CR propagation parameters will change
our results.  However, we cannot prove this assertion without a detailed
reanalysis of the AMS-02 data, beyond the scope of the current paper.

\begin{figure*}
  \hspace*{\stretch{1}}
  \includegraphics[keepaspectratio,width=0.75\textwidth]{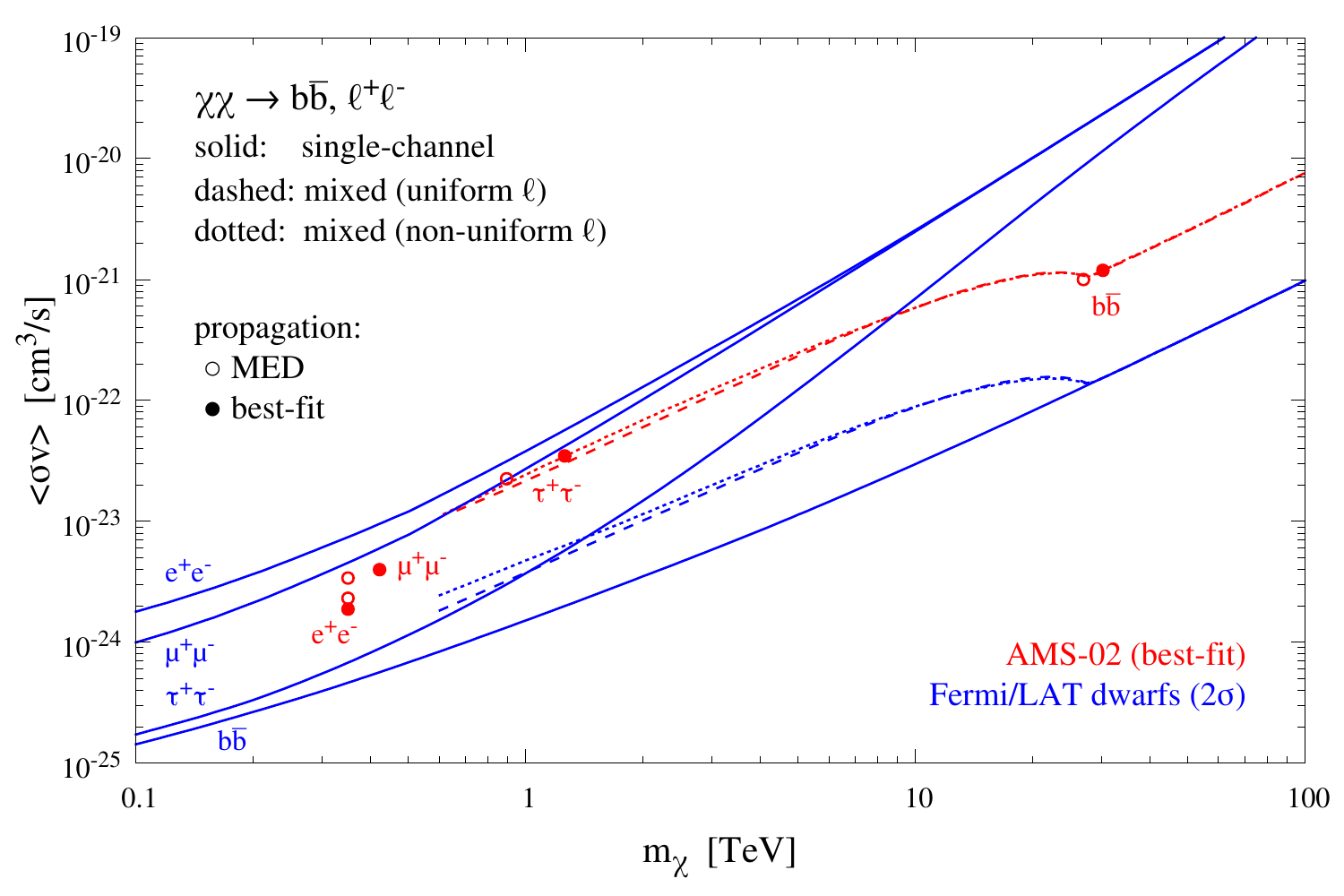}
  \hspace*{\stretch{1}}
  \caption{\label{fig:2b2l}
    AMS-02 best-fit parameters and Fermi/LAT constraints on $\sigmav$
    and $\mchi$ for the $b\bar{b}$ and $\ell\bar{\ell}$ channels.
    Best-fit AMS-02 values for single annihilation channels are shown
    as red circles (taken from Ref.~\cite{Boudaud:2014dta}).  Empty
    circles are for the MED propagation parameters and filled circles
    are for the best-fit propagation parameters.  Fermi/LAT
    constraints for each of these single channels are shown as solid
    blue lines.  The red and blue dashed curves represent respectively
    the AMS-02 best-fit and Fermi/LAT constraint on $\sigmav$ at each
    mass for the mixed $b\bar{b}$+$\ell\bar{\ell}$ channels, assuming
    a common (uniform) branching ratio (BR) into each of the three
    leptons; Fermi/LAT constraints are generated assuming the AMS-02
    best-fit BRs at each mass.  The dotted curves are the same, but
    relaxing the assumption of uniform lepton BRs.  We note that the
    MED propagation parameters are used for the mixed channels.  While
    we do not believe that changing the CR propagation parameters will
    significantly affect our results, proving this assertion is beyond
    the scope of this paper.
  }
\end{figure*}

Figure~\ref{fig:2b2l} shows the AMS-02 best fits and Fermi/LAT dwarf
constraints on $\sigmav$ and $\mchi$ for annihilations into single
channel $\ell\bar{\ell}$ or $b\bar{b}$, as well as mixed cases
$\ell\bar{\ell} + b\bar{b}$.  Note that Figure~\ref{fig:2b2l} is over
a broader mass range than Figure~\ref{fig:2b4tau}.  First let us
discuss the single channel cases.  The Fermi/LAT upper bounds on the
single channel cases are shown as solid blue lines, from bottom to
top: $b\bar{b}$, $\tau^{+}\tau^{-}$, $\mu^{+}\mu^{-}$, $e^{+}e^{-}$.
The $e$ and $\mu$ channels have weaker limits as photon production is
suppressed in these cases, here coming only from the FSR.\footnote{%
  The muon produces a photon in $O(1\%)$ of its decays, though this
  process is not accounted for in \texttt{PYTHIA}.}
The $b$ and $\tau$ channels produce photons through unsuppressed
shower/decay processes; the presence of FSR has little impact on the
constraints for these two channels.  The best-fit AMS-02 points for
these annihilation channels are shown by the red circles.  Empty
circles are for the MED propagation parameters and filled circles are
for the best-fit propagation parameters \cite{Boudaud:2014dta}; the DM
parameters do not differ much between the two cases for any of the
channels.  The AMS-02 $b$ and $\tau$ best-fit points are excluded by
the Fermi/LAT constraints, while the $e$ and $\mu$ points are not. The
two lighter lepton channels fail to be excluded by Fermi/LAT because
they provide positrons in abundance for the AMS-02 signal, while
photon production is suppressed, leading to little expected
sensitivity to these channels via $\gamma$-ray searches.  However,
both the $e$ and $\mu$ annihilation channels, while capable of
producing substantial numbers of positrons, are simply a poor fit to
the AMS-02 spectrum.  Thus, none of these four channels can provide a
reasonable fit to both the AMS-02 and Fermi/LAT results and no single
annihilation channel into quarks, leptons, or gauge bosons can
simultaneously explain the AMS-02 data while remaining in agreement with
Fermi-LAT bounds.  We note that both the MED and the best-fit CR
propagation parameters have been used in studying these single channel
cases.

Reference~\cite{Boudaud:2014dta} also considered the mixed-channel
case, where the DM annihilates into some combination of $b\bar{b}$ and
the three leptons.  Here the MED propagation model is used for the
mixed channels.  Two possibilities were considered: one where all
three leptons were assumed to have a common (uniform) branching ratio
(BR) and one where this assumption is relaxed.  The AMS-02 best-fit
$\sigmav$ as a function of mass are shown as red dashed and dotted
curves for the uniform and non-uniform lepton cases, respectively
(taken from Figures~7 \&~5 in their paper).  Their results were
presented only down to DM masses of 0.6~TeV, hence the termination of
the curves at that mass.  For the uniform case, the leading
annihilation channel is always to $b\bar{b}$, while for the
non-uniform case, $\tau^+\tau^-$ dominates below 20~GeV and $b\bar{b}$
dominates above. The corresponding Fermi/LAT 2$\sigma$ upper limits,
assuming the best-fit BRs, are shown in blue dashed and dotted curves.
The AMS-02 best-fit $\sigmav$ and BRs are strongly excluded by the
Fermi/LAT results.  This does not rigorously imply the AMS-02 and
Fermi/LAT results are in strong conflict for all multi-channel cases,
as the Fermi/LAT constraints will vary if the BRs are allowed to
deviate from their best-fit values.  However, bringing the two results
into compatibility will require the $b\bar{b}$ channel to be heavily
suppressed, as well as the $\tau^{+}\tau^{-}$ channel at the lighter
end of the mass range, which is quite different from the best-fit
case, where $b\bar{b}$ dominates at higher masses and $b\bar{b}$ +
$\tau^{+}\tau^{-}$ accounts for $\ge50$\% of the annihilations at
lower masses.

\begin{figure*}
  \hspace*{\stretch{1}}
  \includegraphics[keepaspectratio,width=0.75\textwidth]{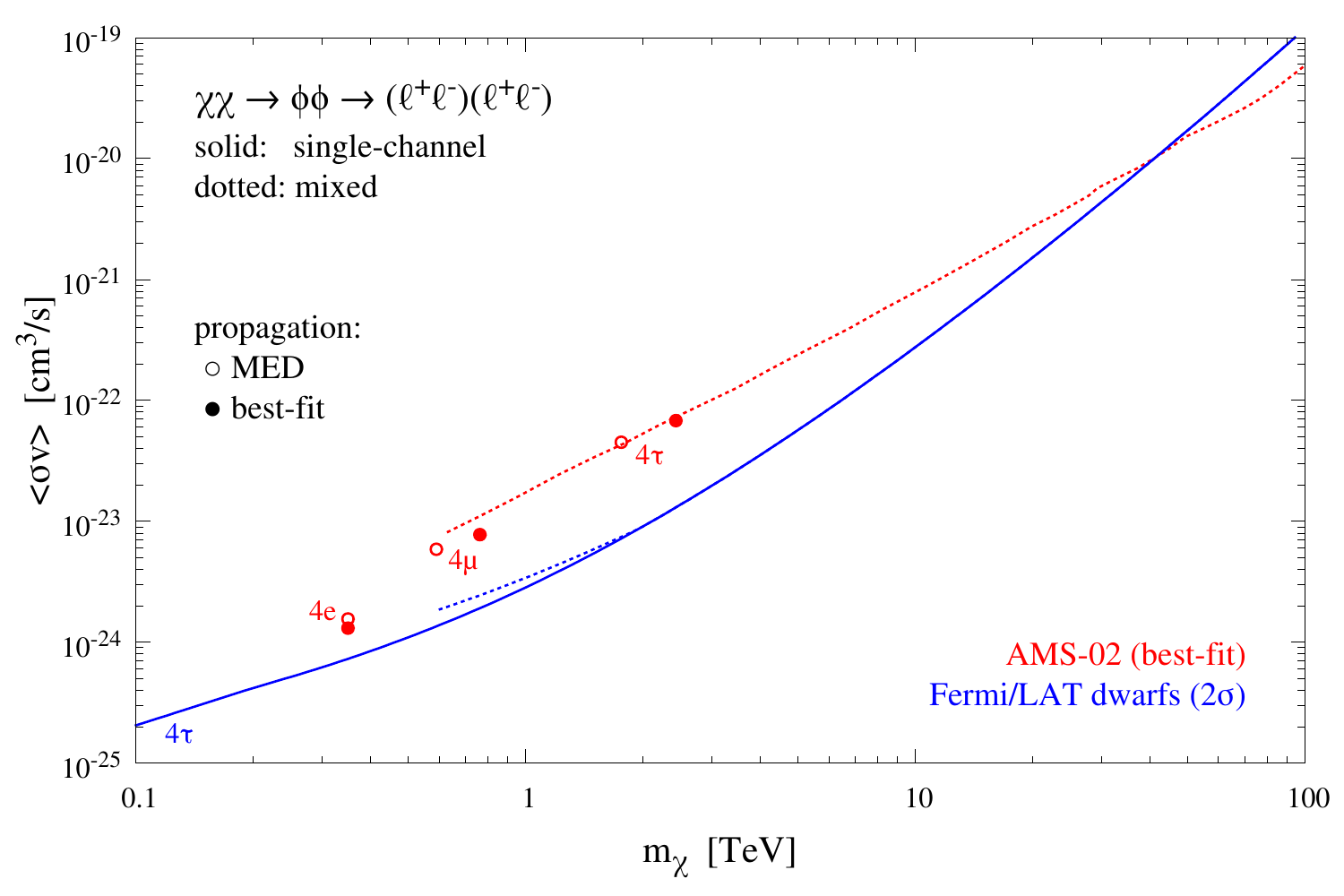}
  \hspace*{\stretch{1}}
  \caption{\label{fig:4l}
    Same as Figure~\ref{fig:2b2l}, but for the mediated 4-lepton channels.
    The solid blue line is the Fermi/LAT bound for the 4-$\tau$ case,
    which excludes the best-fit AMS-02 values.
    The 4-$e$ and 4-$\mu$ Fermi/LAT bounds have not been calculated here,
    but are likely to be weaker than their corresponding best-fit AMS-02
    parameters (see the text).
    Whereas the 4-$e$ case does not provide a good fit to the AMS-02
    data, the 4-$\mu$ case does and thus survives as a viable alternative.
  }
\end{figure*}

Figure~\ref{fig:4l} examines the mediated 4-lepton case with
$\chi\chi\to\phi\phi$ and $\phi\to\ell\bar{\ell}$.  The 4-$\tau$
Fermi/LAT upper limit is shown by the solid blue curve.  Due to the
complexity of implementing their non-leading-order photon production
mechanisms, we have foregone the calculation of limits for the 4-$e$
and 4-$\mu$ cases.  However, as with the 2-lepton cases shown in
Figure~\ref{fig:2b2l}, these $e$ and $\mu$ limits can be expected to
be 1--2 orders of magnitude weaker than the $\tau$ limit.  The AMS-02
best-fit points are again shown by red circles, determined using both
MED propagation (empty circles) and the best-fit propagation
parameters (solid circles) \cite{Boudaud:2014dta}.  The $\tau$ point
is excluded by the Fermi/LAT results.  Without a calculation of the
$e$ and $\mu$ Fermi/LAT constraints, compatibility of the two
experimental results cannot be checked, though the Fermi/LAT
constraints are almost certainly too weak to exclude these two points.
The $e$ channel is a poor fit to the AMS-02 data and thus of little
interest.  However, the 4-$\mu$ channel both provides a good fit to
the AMS-02 data and escapes Fermi/LAT constraints, thus remaining a
viable model.

Reference~\cite{Boudaud:2014dta} considered the mixed 4-lepton case,
with the best-fit $\sigmav$ shown by the dotted red curve in
Figure~\ref{fig:4l}, using the MED propagation model.  The best-fit
BRs are annihilation exclusively to taus for DM masses above 2~TeV,
and a mix of electrons and taus below that, though still dominated by
the tau channel for the masses shown.  The Fermi/LAT constraint for
these best-fit BRs is shown by the dotted blue curve, which becomes
identical to the single-channel 4-$\tau$ constraint above 2~TeV.  The
AMS-02 best-fit $\sigmav$ and BRs are incompatible with the Fermi/LAT
data except at very high masses ($\mchi>40$~TeV).  The analysis shown
in Figure~\ref{fig:4l} includes the case featured by
Ref.~\cite{Boudaud:2014dta} as their favored scenario: a combination
between the four-tau (75\%) and four-electron (25\%) channels for a DM
mass between 0.5 and 1 TeV.  This case is ruled out by Fermi/LAT as an
explanation of AMS-02 data (for $\mchi< 40$ TeV).  The caveats discussed
for the $b\bar{b}$+$\ell\bar{\ell}$ case previously apply here:
allowing the BRs to vary from their best-fit values will change the
Fermi/LAT constraints and could potentially bring these experimental
results into line for this model.  The $\tau$ channel would need to be
strongly suppressed for that to occur, which implies BRs far different
than the best-fit case.  An example where this happens is the
predominantly 4-$\mu$ case which remains a viable model.

In addition to bounds from $\gamma$-rays and antiprotons already
discussed, there are other observational constraints on a DM
interpretation of the $e^+$ excess seen in AMS-02.  DM annihilation
products also include neutrinos.  References
\cite{Sandick:2009bi,Spolyar:2009kx,Erkoca:2010vk} pointed out that a
leptophilic explanation of a positron excess should also produce large
numbers of neutrinos detectable in the IceCube neutrino observatory.
Currently IceCube upper limits are in tension with the DM bestfit to
the positron anomaly, assuming MED propagation parameters, when DM
annihilates into $W^+ W^-$ \cite{Aartsen:2013dxa}.  In addition,
measurements of the CMB temperature and polarization provide
constraints on the annihilation cross section of DM
\cite{Galli:2009zc,Giesen:2012rp,Slatyer:2009yq,Finkbeiner:2011dx,
Natarajan:2012ry,Cline:2013fm}.  In the near future, papers from the
Planck Satellite with extremely strong bounds are expected and could
rule out the DM scenario completely.

In summary, we have used the Fermi/LAT dwarf galaxy data in order to
constrain dark matter as an explanation of the positron excess seen in
HEAT, PAMELA, and AMS-02.  In particular, this paper has focused on the
annihilation channels that best fit the current AMS-02 data
\cite{Boudaud:2014dta}.  We first considered the single channel case
of DM annihilating directly to $b\bar{b}$ and leptons
$\ell\bar{\ell}$, as well as to 4-leptons (via a mediator $\phi$); we
then considered the multi-channel case where annihilation proceeds
through a combination of channels.  We used $b \bar{b}$ as our proxy
for all other quarks, gauge bosons, and Higgs bosons, since the
spectra and amplitudes are similar, but considered each lepton flavor
separately.  For the single channel case, we found that dark matter
annihilation into $\{b\bar{b},e^+e^-,\mu^+\mu^-,\tau^+\tau^-,4$-$e,$
or $4$-$\tau \}$ cannot both provide a good fit to AMS-02 and avoid
the 2$\sigma$ upper limit from Fermi/LAT. The AMS-02 best-fit BRs and
$\sigmav$ used for the analysis of this paper assume either the MED
propagation model defined in Refs.~\cite{Donato:2003xg,Maurin:2001sj}
or the best-fit propagation parameters
\cite{Boudaud:2014dta}. Multi-channel annihilations are also highly
constrained by Fermi/LAT's measurement of $\gamma$-ray flux. We use
the MED propagation model parameters for the mixed channel cases, but
doubt that other reasonable choices of CR propagation parameters would
change our results.  Specifically, we find that the Fermi/LAT
2$\sigma$ upper limits, assuming the best-fit AMS-02 BRs, exclude the
annihilation into a combination of $b\bar{b}$ and the three leptons
for DM masses 600~GeV $\leq \mchi \leq$ 100~TeV. In addition, the
Fermi/LAT upper limit is incompatible with the AMS-02 best fit
$\sigmav$ and BRs for annihilation into a mix of the mediator driven
4-lepton channels except for very high masses ($\mchi\ge
40$~TeV). However, this does not rigorously imply that the results
from AMS-02 and Fermi/LAT are in strong conflict for all multi-channel
cases, as the Fermi/LAT constraints will vary if the BRs are allowed
to deviate from their best fit AMS-02 values. In order to reconcile
both experiments, the BRs considered would have to deviate
significantly from their best-fit values. We find that the dark matter
annhiliation into 4-$\mu$ provides a good fit to the AMS-02 data and
escapes Fermi/LAT constraints. The reason for the 4-$\mu$ channel
escaping Fermi/LAT upper limits is that it provides positrons in
abundance for the AMS-02 signal, while photon production is
suppressed. Hence, for the best-fit BR to the AMS-02 data, we find
only one viable DM annihilation channel that survives Fermi/LAT
constraints and provides a good fit to the current AMS-02 data: the
4-$\mu$ channel.
 
We briefly mention ways around these conclusions.  First, for the
mixed channel cases we have only studied the best fits to AMS-02 data
(found in Ref.~\cite{Boudaud:2014dta}). It is possible that there are
other BRs that still provide reasonably good fits (though not the best
fits) to the AMS-02 data that are not ruled out by Fermi/LAT bounds from
dwarf spheroidals.  A joint statistical analysis of the AMS-02 and
Fermi/LAT would be required to check for other alternatives.  Second,
although it is extremely unlikely that the boost factor required by
the AMS-02 data is due to a nearby clump of DM that is 100--1000 times
as dense as its surroundings, perhaps part of the boost factor is due
to a clump of, say, a factor of 10 (again, unlikely).  In that case
the required annihilation cross-section to explain the AMS-02 data could
be lower, and more channels would remain viable.  Third, it is
possible that part of the AMS-02 signal is due to pulsars, and part due
to DM annihilation.  Again, more DM annihilation channels might then
still remain compatible with Fermi/LAT bounds from dwarf spheroidals.
These latter two caveats would also apply to bounds on these scenarios
from the CMB including those expected from upcoming Planck data.

\section*{Acknowledgments}
A.L.\ acknowledges support from the DOE under grant DOE-FG02-95ER40899
and from the Michigan Center for Theoretical Physics.  A.L.\ and
D.S.\ are grateful for financial support from the Swedish Research
Council (VR) through the Oskar Klein Centre.

\end{document}